\documentclass[%
preprint,
 amsmath,amssymb,
 aip,
floatfix,
]{revtex4-2}

\usepackage{graphicx}
\usepackage{dcolumn}
\usepackage{bm}

\usepackage[utf8]{inputenc}
\usepackage[T1]{fontenc}
\usepackage{mathptmx}
\usepackage{comment}
\usepackage{color}
\usepackage[dvipsnames]{xcolor}
\usepackage{soul}
\usepackage{xspace}
\usepackage{xcolor}
\usepackage{hyperref}
\usepackage{cleveref}

\usepackage{siunitx}[=v2]
\ifdefined\qty\else
\ifdefined\NewCommandCopy
\NewCommandCopy\qty\SI
\NewCommandCopy\qtyproduct\SI
\NewCommandCopy\qtyrange\SIrange
\else
\NewDocumentCommand\qty{O{}mm}{\SI[#1]{#2}{#3}}
\NewDocumentCommand\qtyproduct{O{}mm}{\SI[#1]{#2}{#3}}
\NewDocumentCommand\qtyrange{O{}mm}{\SIrange[#1]{#2}{#3}}
\fi
\fi
\ifdefined\unit\else
\ifdefined\NewCommandCopy
\NewCommandCopy\unit\si
\else
\NewDocumentCommand\unit{O{}m}{\si[#1]{#2}}
\fi
\fi

{%

\newcommand{\methods}{\textbf{Methods}\xspace}

\newcommand{\hA}{\hat{A}}

\begin{document}

\baselineskip24pt

\section*{Government License}

The submitted manuscript has been created by UChicago Argonne, LLC,
Operator of Argonne National Laboratory (“Argonne”). Argonne, a
U.S. Department of Energy Office of Science laboratory, is operated
under Contract No. DE-AC02-06CH11357. The U.S. Government retains for
itself, and others acting on its behalf, a paid-up nonexclusive,
irrevocable worldwide license in said article to reproduce, prepare
derivative works, distribute copies to the public, and perform
publicly and display publicly, by or on behalf of the Government.  The
Department of Energy will provide public access to these results of
federally sponsored research in accordance with the DOE Public Access
Plan. http://energy.gov/downloads/doe-public-access-plan.

\newpage

\title{Demonstration of an AI-driven workflow for autonomous high-resolution scanning microscopy}

\author{Saugat Kandel}%
\affiliation{ 
Advanced Photon Source, Argonne National Laboratory, Lemont, IL 60439.
}%

\author{Tao Zhou}%
\affiliation{ 
Nanoscience and Technology Division, Argonne National Laboratory, Lemont, IL 60439.
}%

\author{Anakha V Babu}%
\affiliation{ 
Advanced Photon Source, Argonne National Laboratory, Lemont, IL 60439.
}%

\author{Zichao Di}%
\affiliation{ 
Mathematics and Computer Science, Argonne National Laboratory, Lemont, IL 60439.
}%

\author{Xinxin Li}%
\affiliation{ 
Nanoscience and Technology Division, Argonne National Laboratory, Lemont, IL 60439.
}
\affiliation{Consortium for Advanced Science and Engineering, University of Chicago, Chicago, Illinois 60637, USA}%

\author{Xuedan Ma}%
\affiliation{ 
Nanoscience and Technology Division, Argonne National Laboratory, Lemont, IL 60439.
}
\affiliation{Consortium for Advanced Science and Engineering, University of Chicago, Chicago, Illinois 60637, USA}%

\author{Martin Holt}%
\affiliation{ 
Nanoscience and Technology Division, Argonne National Laboratory, Lemont, IL 60439.
}%

\author{Antonino Miceli}%
\affiliation{ 
Advanced Photon Source, Argonne National Laboratory, Lemont, IL 60439.
}%

\author{Charudatta Phatak}
\affiliation{%
Materials Science Division, Argonne National Laboratory, Lemont, IL 60439.
}%

\author{Mathew Cherukara}
\email{mcherukara@anl.gov, skandel@anl.gov}
\affiliation{ 
Advanced Photon Source, Argonne National Laboratory, Lemont, IL 60439.
}

\date{\today}

\begin{abstract}

 With the continuing advances in scientific instrumentation, scanning microscopes are now able to image physical systems with up to sub-atomic-level spatial resolutions and sub-picosecond time resolutions. Commensurately, they are generating ever-increasing volumes of data, storing and analysis of which is becoming an increasingly difficult prospect. One approach to address this challenge is through self-driving experimentation techniques that can actively analyze the data being collected and use this information to make on-the-fly measurement choices, such that the data collected is sparse but representative of the sample and sufficiently informative. Here, we report the Fast Autonomous Scanning Toolkit (FAST) that combines a trained neural network, a route optimization technique, and efficient hardware control methods to enable a self-driving scanning microscopy experiment. The key features of our method are that: it does not require any prior information about the sample, it has a very low computational cost, and that it uses generic hardware controls with minimal experiment-specific wrapping. We test this toolkit in numerical experiments and a scanning dark-field x-ray microscopy experiment of a WSe\textsubscript{2} thin film, where our experiments show that a FAST scan of <25\% of the sample is sufficient to produce both a high-fidelity image and a quantitative analysis of the surface distortions in the sample. We show that FAST can autonomously identify all features of interest in the sample while significantly reducing the scan time, the volume of data acquired, and dose on the sample. The FAST toolkit is easy to apply for any scanning microscopy modalities and  we anticipate adoption of this technique will empower broader multi-level studies of the evolution of physical phenomena with respect to time, temperature, or other experimental parameters.

\end{abstract}

\maketitle

\section{Introduction}

Scanning microscopes are versatile instruments that use photons, electrons, ions, neutrons, or mechanical probes to interrogate atomic-scale composition, topography, and functionality of materials, with up to sub-atomic spatial resolution and sub-picosecond time resolution~\cite{goldstein_scanning_2018,zuo_advanced_2017, voigtlander_scanning_2015}. Notwithstanding the variation in the probe modalities, these instruments all rely on a scan of the sample to generate spatially resolved signals that are then collected to form an image of the sample. Ongoing advances in instrumentation, such as the development of next-generation x-ray and electron detectors\cite{Hiraki:a60117,Tate2016}, has meant that scanning microscopes can now image faster, and at higher resolutions, than ever before. We can now envision a broad use of these instruments to study not only static systems, but also multi-level studies of dynamic evolution of materials with time, temperature, or other parameters, even \textit{in situ} or \textit{operando}~\cite{kalinin_arxiv_2022}. Fine-resolution large-field-of-view scanning experiments, however, come with some significant drawbacks: the volume of data generated and the probe-induced damage to the sample can be prohibitively large. For example, it is now routinely possible to perform x-ray imaging of \qty{1}{mm^3} volumes at \qty{\approx 10}{\nm} resolution, but this generates $\approx 10^{15}$ voxels of data~\cite{PSI-chip, Jiang:2021jw} and requires a commensurately high probe dose~\cite{du_jac_2021}. Meanwhile, the information of interest in these experiments is often concentrated in sparse regions that contain interfaces, defects, or other specific structural elements. Directing the scan to only these locations could greatly reduce the scan time and data volume, but it is difficult to obtain this information \textit{a priori}. Addressing this challenge with a human-in-the-loop protocol, where an experienced user examines the data acquired to identify trends and guide the scan, can be tedious and prohibitively time consuming (in comparison to the experimental acquisition time). Given these factors, the development of autonomous acquisition techniques that can continuously analyze acquired data and drive the sampling specifically towards regions of interest is imperative so as to make full use of the potential of these scientific instruments.

In parallel to the advances in scientific instrumentation, the last decade has also seen the rapid development of deep learning (DL) techniques and their applications in all domains of science and technology, including for the acceleration and enhancement of advanced microscopy methods~\cite{cherukara_apl_2020,chan_apr_2021,yao_npjcm_2022,babu_arxiv_2022}. These DL-based inversion methods are enabling real-time data analysis, which is in turn opening the door to self-driving techniques that make real-time acquisition decisions based on the real-time data streams. Such self-driving or autonomous experimentation methods~\cite{hase_tc_2019} are methods that combine automated experimental control with on-the-fly data-driven decision making so that an algorithm adaptively explores parameter spaces of interest and conducts new experiments until it achieves a pre-defined completion criterion \cite{burger_nature_2020}. These methods therefore have the potential to not only remove the need for constant human supervision and intervention in experiments, but also  make optimal choices in parameter spaces that are too large for humans to easily contextualize. As such, they have the potential to revolutionize experimental design in many scientific fields including the field of imaging and materials characterization. 

In general, the use of data-driven priors to direct future experiments is a Bayesian search problem, for which the use of off-the-shelf deep learning methods usually do not suffice \cite{vasudevan_npjcm_2021}. Specific to microscopy, a popular Bayesian search approach  is to use unsupervised (without pre-training) Gaussian Processes (GPs) that could continuously determine the spatial locations that we are most uncertain about, then direct the scanning to these locations \cite{noack_scirep_2019,noack_scirep_2020,noack_natrevphys_2021,vasudevan_acsnano_2021,kalinin_acsnano_2021,garnett_bayesoptbook_2023}. While GPs are powerful techniques, their computational cost tends to scale \textit{cubically} with the number of points acquired. The decision making time increases during the experiment and quickly exceeds the acquisition time for the measurement itself. The development of scalable GPs is a significant area of research, but these methods are not yet ready for application in large-scale imaging problems~\cite{liu_itnnls_2020}. General supervised alternatives  such as reinforcement learning can be powerful and fast, but they often require costly pre-training and tend to ignore the global state of the parameter space in exchange for a local search; as such they have only found limited traction for scanning imaging modalities \cite{schloz_arxiv_2022}.

Specifically for scanning microscopy applications, Godaliyadda \textit{et al.}~\cite{godaliyadda_ei_2016}  have proposed to achieve computationally efficient autonomous sampling with the Supervised Learning Approach for Dynamic Sampling (SLADS) technique. The SLADS technique uses curated feature maps to quantify the current measurement state and predict the total image quality improvement obtained by measuring a given point, thereby informing the choice of which point to measure next. Variations of this technique have found applications in live steering for dose-efficient crystal positioning for crystallography~\cite{scarborough_jsr_2017}, and for imaging with transmission electron microscopy ~\cite{hujsak_micron_2018} and mass spectrometry~\cite{hu_acsmsa_2022} methods. These works, however, either involve training with and reconstruction of binary images only~\cite{scarborough_jsr_2017,hujsak_micron_2018}, or, require extensive training with images closely related to the sample under study~\cite{hu_acsmsa_2022}. As such, they are difficult to translate to imaging settings with more complex images, particularly for imaging without any prior assumptions about the sample. Meanwhile, Zhang \textit{et al.}~\cite{zhang_ei_2018} have incorporated a neural network (NN) within the SLADS method (for the SLADS-Net method) and shown in numerical experiments that it is sufficient to train the method on only a generic image, eschewing any prior knowledge about the sample, to produce high-fidelity image with sparse sampling. However, this has not yet been demonstrated in experiment.

In this work, we report the \textbf{F}ast \textbf{A}utonomous \textbf{S}canning \textbf{T}oolkit (FAST) that combines the SLADS-Net method, a route optimization technique, and efficient and modular hardware controls  to make on-the-fly sampling and scan path choices for synchrotron-based scanning microscopy.
This method relies on sample-agnostic training to dynamically measure and reconstruct a complicated (non-binary) sample, distinguishing this toolkit from existing SLADS-based workflows. Moreover, its computational cost is negligible compared to the acquisition time even when run on a low-power edge computing device placed at a synchrotron beamline, which presents a significant advantage over more generic autonomous experimentation techniques. These characteristics enable the application of our workflow in the high-precision nanoscale scanning x-ray microscopy instrument present at the hard x-ray nanoprobe beamline at the Advanced Photon Source.

We validate the FAST scheme through real time demonstration at the hard x-ray nanoprobe beamline at the APS \cite{Winarski:ie5086}. A few-layer exfoliated two-dimensional WSe$_2$ thin film was chosen as a representative example; the preparation process for the thin film often leaves microscopic air bubbles trapped underneath the thin film, deforming the 2D material. We show that an adaptive scan of $<25\%$ of the sample is sufficient to produce a high-fidelity reconstruction that identifies all the bubbles within the field of view,  and even to acquire quantitative information about the film curvature induced by these bubbles. The scheme quickly identifies the deformed part of the 2D material and focuses its attention there, while ignoring regions of the film that are flat and homogeneous. Film curvature reconstructed from the adaptive scan ($<25\%$ coverage) is consistent with that reconstructed from full-grid scan ($100\%$ coverage). Given these characteristics, the FAST scheme can be directly applied in other scanning techniques and instruments at the APS and elsewhere, and may underpin the development of many multi-level experimental studies.

\section{Results}\label{sec:results}
\begin{figure}
\centering
\includegraphics[width=0.85\textwidth]{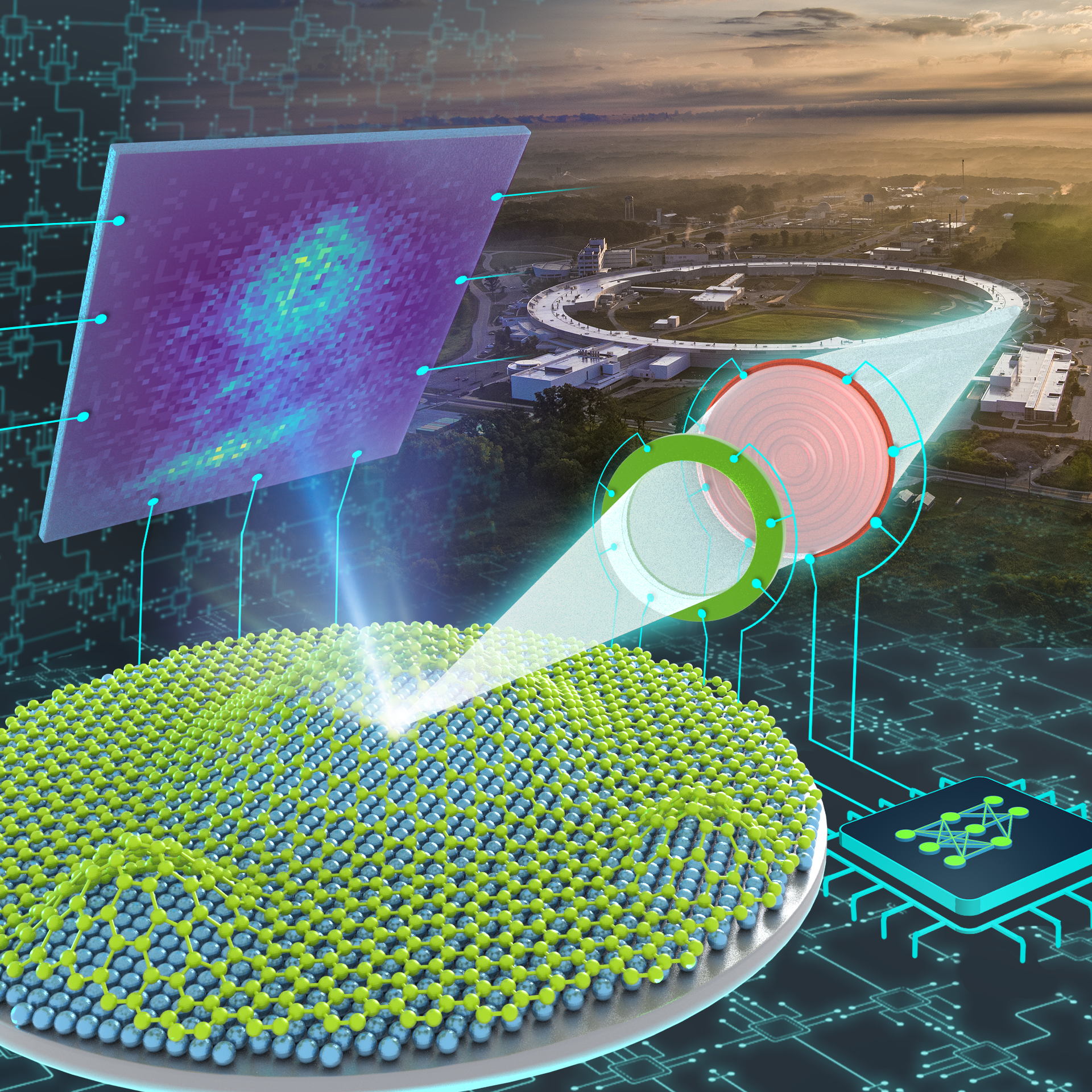}
\caption{(Artist's representation) The APS synchrotron produces a coherent x-ray beam that is focused using a zone plate setup. It strikes a WSe\textsubscript{2} film (green) exfoliated onto a Si substrate (blue), which generates diffraction patterns that are collected by a two-dimensional detector. Above the bubbles, the lattice of the film rotates, shifting the diffracted intensities away from its nominal positions. The beam position as well as the detector acquisition are autonomously controlled by the FAST AI-based workflow.}\label{fig:artist}
\end{figure}

~\Cref{fig:artist} shows the experimental setup that scans a focused x-ray beam on a sample while acquiring a two-dimensional diffraction image at each point. The live demonstration was performed on a few-layer WSe\textsubscript{2} sample with the detector placed along the 008 Bragg peak, with $2\theta = \ang{43.1}$ at \qty{10.4}{\keV}. The diffraction patterns were processed on the detector computer (see \methods) to generate the integrated intensities for use in the FAST workflow. The final output of the workflow is a dark-field image of the WSe\textsubscript{2} sample. 



\subsection{Self-driving scanning microscopy workflow}\label{sec:workflow}
\begin{figure}
\centering
\includegraphics[width=0.99\textwidth]{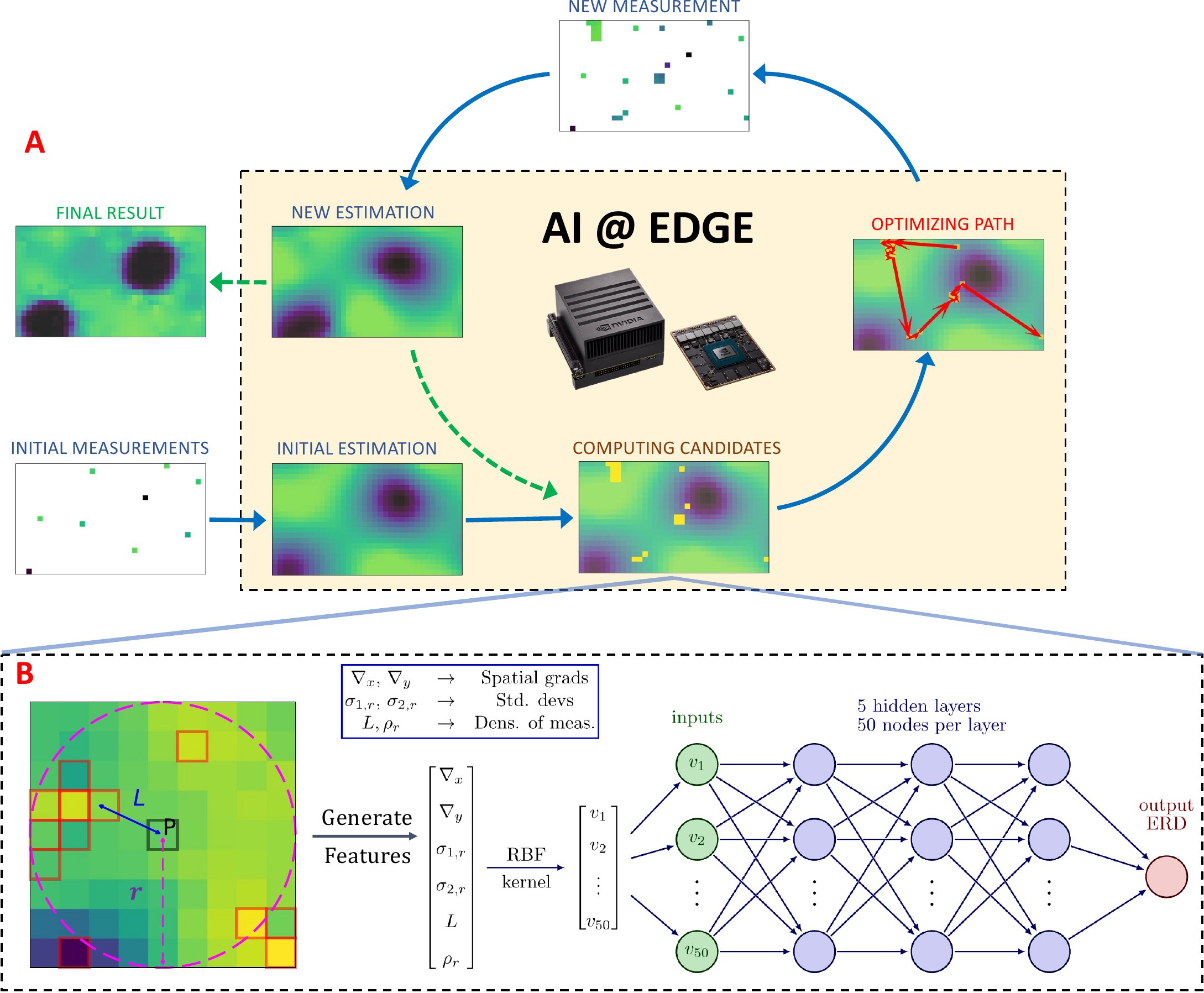}
\caption{The FAST workflow: (A) A set of random initial measurements are transferred to the edge device which sequentially generates an initial sample estimate, computes the candidate points to be measured next, and calculates the travel path for the measurement. The new measurements are combined with the existing measurements and used to calculate a new estimate, and the process is repeated until it achieves a completion criterion. (B) The candidate computation starts by examining the local neighborhood (with radius $r$) of each unmeasured point $P$, with the highlighted points indicating points already measured, to generate a 6-dimensional feature vector. The feature vector is transformed to a 50-dimensional vector using the Radial Basis Function (RBF) kernel and used as input to a multi-layer NN. The NN then predicts the expected improvement in the image (ERD) from measuring the point $P$. A set of unmeasured pixels with the highest ERD are selected as candidates for the next measurement.}\label{fig:workflow}
\end{figure}

\Cref{fig:workflow}A broadly illustrates the FAST workflow for the experiments reported here. To initiate the workflow, a low-discrepancy quasi-random selection (generated using the Hammersely sequence~\cite{wong_jgt_1997}) of sample position is measured corresponding to 1\% of the total area of interest. 
The integrated intensities of the measurements are transferred to the edge device, an NVIDIA Jetson Xavier AGX \cite{AGX} located adjacent to the detector, which used Inverse Distance Weighted (IDW) interpolation to estimate the dark-field image. The estimated image serves as input for the decision-making step whereby the prospective measurement points are identified.

This self-driving workflow adopts the Supervised Learning Approach for Dynamic Sampling using Deep Neural Networks (SLADS-Net) algorithm \cite{zhang_ei_2018} to find the prospective measurement points. In effect, the SLADS-Net algorithm uses the current measurements to identify the best unmeasured points that, when added to the existing dataset, would have the greatest effect on the quality of the reconstructed image. As illustrated in \Cref{fig:workflow}B, this is accomplished by, first, representing each unmeasured point as a feature vector with elements that depend on the measurement state in the neighborhood of the point. These feature vectors are used as input for a pre-trained multi-layer perceptron. The neural network then predicts the expected reduction in distortion (ERD), a metric (loosely speaking) for the expected improvement in the reconstruction quality obtained from measuring this unmeasured point, individually for each unmeasured point. The original SLADS-Net algorithm simply uses the unmeasured point with the highest ERD for the next measurement, and repeats this procedure pointwise. In practice, if the measurement procedure and the motor movements are fast, then the ERD calculation also has to be commensurately fast to reduce the dead-time in the experiment. In this work, we mitigate this requirement by instead selecting a batch of points that have the highest ERD, sorted in descending order---we found that a batch of 50 points adequately minimized the experimental dead-time while still ensuring that the overall measurement was adequately sparse.

The coordinates of these 50 points are passed on to a route optimization algorithm, based on Google's OR-Tools \cite{ortools}, to generate the shortest path for the motors to visit all of the them. This path is appended to the look-up table in the EPICS \cite{EPICS} scan record, which then kicks off the data acquisition. Henceforth, the scan is automatically paused after every 50 points, raising a flag which event triggers a callback function on the edge device. There, a new estimated dark field image of the sample is generated, and the coordinates for the next 50 prospective points are computed. The scan is resumed after the EPICS scan record receives the new coordinates for the optimized scanning path. The actual scanning of the focused x-ray beam is achieved by moving two piezoelectric linear translation motors in step mode. The detector exposure time is set to \qty{0.5}{\s} and comes with an overhead  of \qty{0.2}{\s}.

For the \qtyproduct{200 x 40}{} pixels object described in Section \ref{subsec:experiment}, the workflow required \qty{\approx 0.15}{\second} to compute the new positions, \qty{\approx 42}{\second} to scan the set of 50 positions, and a total of \qty{\approx 0.37}{\second} to process the diffraction patterns and communicate the measurements.  This represents an overhead of $\lessapprox 2\%$. The workflow is currently entirely CPU-bound, relying on the on-board 8-core ARM CPUs, and does not take advantage of the GPU bundled into the NVIDIA AGX device. In the future, we expect to perform the computation in a parallelized and asynchronous fashion, which would further reduce this overhead. These timing results  showcase the rapid data-driven decision-making ability that is characteristic of the FAST workflow.

We also note that, for all the results reported in this work, the underlying NN was trained on a single generic image with no relation to microscopy. For details about the SLADS-Net algorithm and the \textit{sample-agnostic} training procedure, the reader is referred to the \methods section.

\subsection{Numerical demonstration for scanning dark-field microscopy}
\label{subsec:simulation}
\begin{figure}[ht]
	\includegraphics[width=0.99\textwidth]{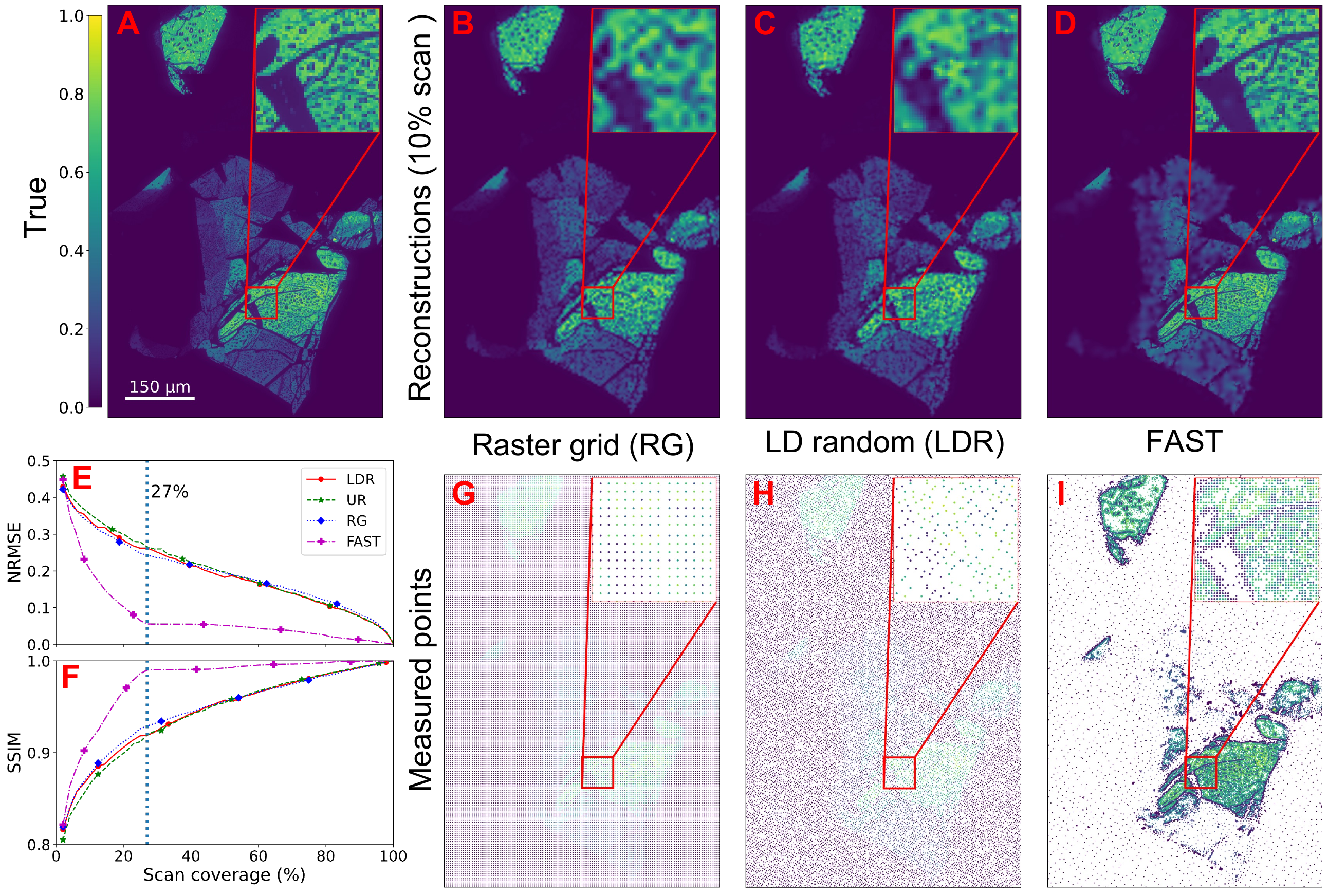}
	\caption{Numerical comparison of sampling methods: (A) shows the ground truth with the color scale representing the normalized intensity, (B-D) show respectively the RG, LDR, and FAST reconstructions at 10\% scan coverage, and (G-I) show the actual scan points that produce these reconstructions. (E-F) show the evolution of the NRMSE (lower is better) and SSIM (higher is better) as a function of the scan coverage. The FAST reconstruction stabilizes at $~27\%$ coverage while the other techniques take significantly longer to reach the same quality. }\label{fig:simulation}
\end{figure}

We first validated the  performance of the proposed workflow through a numerical experiment on a set of pre-acquired dark-field microscopy data. Here, we compared the FAST sampling with three static sampling techniques:
\begin{enumerate}
    \item \textbf{Raster grid (RG)} For a test sampling percentage, we generated a equally  spaced raster grid that provides a uniform coverage of the sample. 
    \item \textbf{Uniform random (UR) sampling} The measurement pixels were drawn from a uniform random distribution.
    \item \textbf{Low-discrepancy (LDR) random sampling} For each measurement percentage, we generated a low-discrepancy sampling grid using the quasi-random Hammersly sequence.
\end{enumerate}
The test dataset is a dark field image of size \qtyproduct{600 x 400}{} pixels which represents 240,000 possible measurement positions. This covers a physical area of \qtyproduct{900x 600}{\micro\m} and encloses multiple flakes of WSe$_2$ with various thicknesses, with the thicker regions associated with regions of higher brightness in the image (\Cref{fig:simulation}). At this spatial resolution, only medium and large sized bubbles (with diameter > 2 um) can be observed. As explained previously, the bubbles deform the surface and shift the Bragg peak of the 2D materials away from their theoretical (flat region) positions, resulting in regions of darker contrast.  Finally, the image also contains flake-free regions that have zero integrated intensities. 

For this comparison, we first initialized the FAST sampling with a 1\% measurement coverage (as described above), then successively measured  50 additional points at iteration. For each FAST measurement, we also generate RG, UR, and LDR measurement masks with the same number of scan points. In this fashion, we generate a sequence of sampling masks and the associated reconstructions until we achieve 100\% sampling.

We present the numerical results in \Cref{fig:simulation}, where we show a comparison of the various methods at 10\% sampling.  Note that while the proposed method internally uses the fast IDW algorithm for the inpainting, the final images presented here are calculated using the higher quality biharmonic inpainting technique \cite{lasiecka_ijmms_2018}.  The uniform random scheme performs worse than the LD-random and raster grid schemes and is not shown in the figure.  In \Cref{fig:simulation}A-D, we can see that the FAST sampling is able to reproduce with high fidelity the flake boundaries, the bubbles, and the regions of transition between the varying levels of thicknesses. In contrast, the LDR and raster schemes produce much lower quality reconstructions of these features. \Cref{fig:simulation}E shows an evolution of the normalized root mean squared error (NRMSE) and \cref{fig:simulation}F the structural similarity metric (SSIM) (which measures multiscale perceptual similarity) for the different sampling techniques.  It is evident that FAST produces high quality reconstructions at much lower measurement percentages than the examined static sampling techniques.  We note that the result could be further improved in the future by using a more sophisticated inpainting technique within the FAST method. To understand how FAST outperforms the other methods under the same sampling condition, we show the actual measured positions of the various schemes at 10\% coverage (\Cref{fig:simulation}G-I). FAST preferentially samples the regions with significant heterogeneity over the homogeneous regions. This is particularly useful for sparse samples, where the time spent sampling from empty regions adds little additional information.

\subsection{Experimental demonstration}
\label{subsec:experiment}
\begin{figure}[ht]
\includegraphics[width=0.99\textwidth]{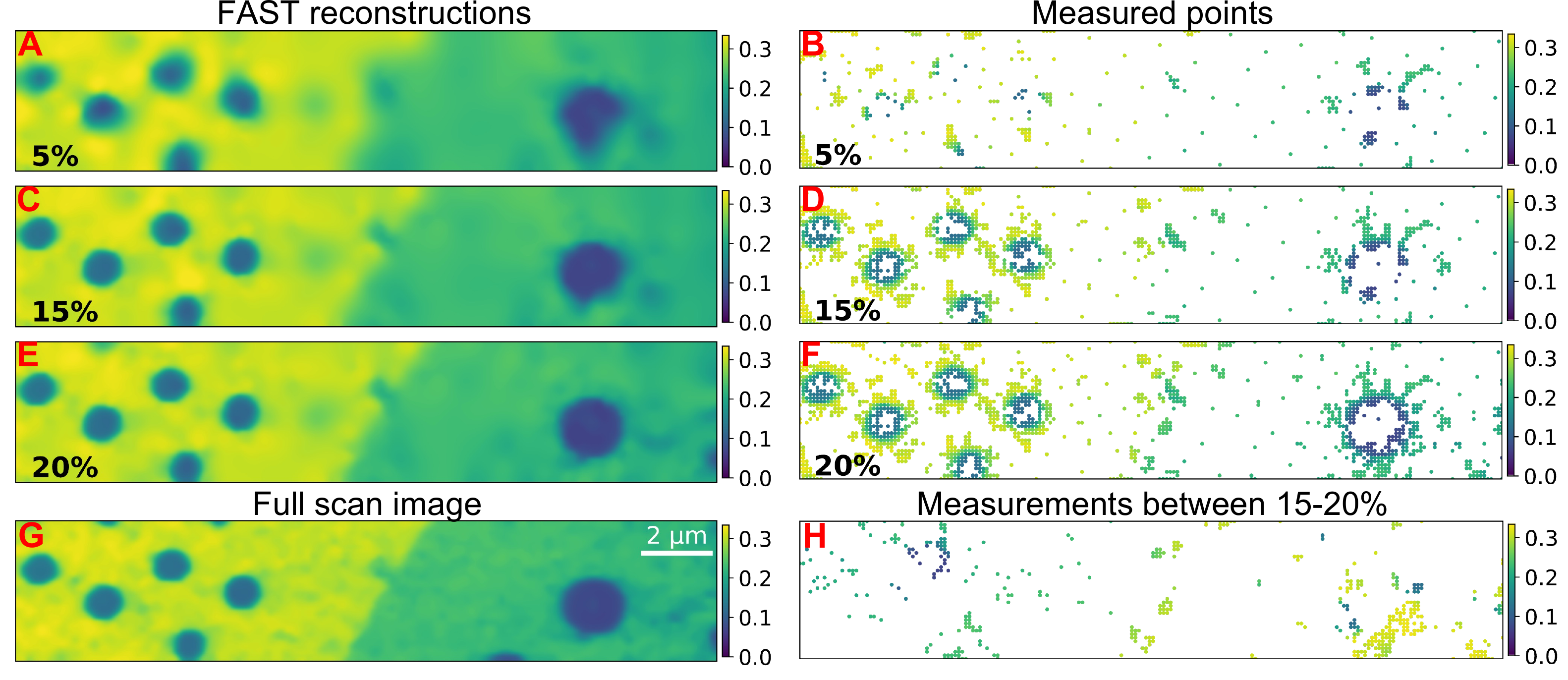}
\caption{Evolution of the FAST scan: (A, C, E) show the reconstruction at 5\%, 15\%, and 20\% reconstructions respectively, (B, D, F) show the corresponding actual measurement points. (G) shows the image obtained through a full-grid pointwise scan. The color scale in (A-G) show the normalized intensities. (H) shows only the points sampled between 15\% and 20\% coverage. }\label{fig:experiment}
\end{figure}

We next demonstrate the application of the FAST workflow in a live experiment at a synchrotron beamline. A video showing the sampling, recorded live during the actual experiment, is available here~\cite{smart_scan_video}. Other than starting the workflow scripts at the beginning, the entire experiment was unmanned and fully automated. In order to measure the deformed WSe$_2$ flakes in details, a higher spatial resolution of 100 nm was chosen. This limits the field of view to \qtyproduct{20 x 4}{\um} for a scan point density of \qtyproduct{200 x 40}{} points. 

In \Cref{fig:experiment} we show the reconstructed dark field image (subplots A,C,E) and the measurement points (subplots B,D,F) from \qtyrange{5}{20}{\%} coverage and compare them to that obtained from raster scanning the sample with 100\% coverage(subplot G).  We see that the FAST method identifies some of the regions of hetereogeneity --- the edges of the bubbles --- and starts to preferentially sample these regions within \qty{5}{\percent} coverage of the sample. At \qty{15}{\percent} coverage, these regions are extensively sampled. The reconstruction does not change significantly between \qtyrange{15}{20}{\percent}, indicating that the reconstruction has stabilized. Moreover, the \qty{20}{\percent} reconstruction also contains sharp and accurate reproductions of all the major features present in the full scan image. 

A point of interest is that the partially scanned bubble at the bottom right corners of \Cref{fig:experiment}E-G shows up only in the 20\% scan, and not in the 15\% scan. To explain this, we note that the  5\% scan, and therefore the initial 1\% random sampling, does not contain any measurements in the neighborhood of this bubble. The FAST scheme favors exploitation of regions it knows to be heterogeneous over exploration of this fully unknown region, and therefore only explores this region much later in the measurement process (\Cref{fig:experiment}H). This is, in fact, an instance of the general exploration-exploitation tradeoff that exists in all Bayesian search procedures~\cite{brochu_arxiv_2010}.  Potential mitigation steps could be to sample more initially (say 5\% random points), or to deliberately introduce diversity into each batch of measurement points.

\begin{figure}[ht]
\includegraphics[width=0.99\textwidth]{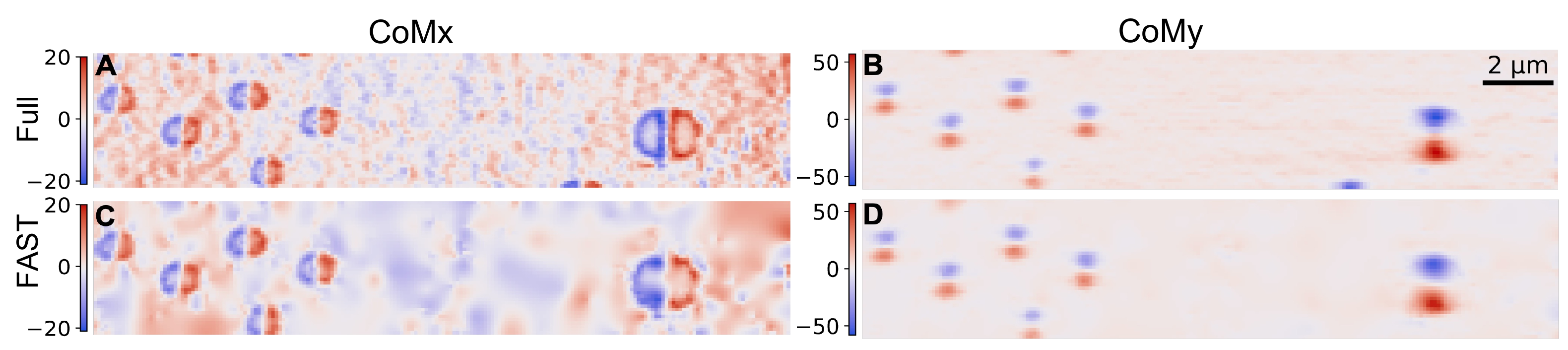}
\caption{Comparison of the per measured point center of mass of the diffraction patterns between the FAST scan at 20\% coverage and full-grid scan. Subplots (A) and (B)  show the inpainted COMx and COMy, respectively, for the full-grid raster FAST scan, and subplots (C) and (D) for the FAST scan.}\label{fig:coms}
\end{figure}

So far we have reduced the diffraction image measured at each point to one single quantity (integrated intensity) in order to guide the automated experiment. These images often need to be reprocessed after the experiment to extract additional physically relevant results. Notably, the intensity distribution in the diffraction patterns contains information about the strain as well as the rotation of the crystal lattice, and in this case, the curvature of the 2D materials due to the bubbles underneath. A simple center of mass calculation in the X direction (CoMx) would yield the magnitude of the film curved in the XZ plane. The curvature (deviation of the CoMx from its nominal value) is the smallest around the center of the bubble and the largest at the edge. It also changes sign going from the left side to the right side. Center of mass calculation in the Y direction yields the magnitude of the film curved in the YZ plane. The results look slightly different from the CoMx calculations due to the way the shifted Bragg peak intersects with the Ewald's sphere. \Cref{fig:coms}A and B shows respectively the CoMx and CoMy obtained from raster scan with 100\% coverage on the area of interest. The unit is the number of pixel shift, relative to the center of the nominal diffraction pattern. \Cref{fig:coms}C and B shows respectively the CoMx and CoMy obtained with FAST. The curvature information of the film were faithfully reproduced despite scanning just 20\% of the entire area. For more information on the reconstruction of the CoM maps, he reader is referred to the \methods section.

%


\section{Discussion}

In this work, we have showcased the FAST workflow that combines a sparse sampling algorithm with route planning to drive a scanning diffraction microscopy experiment at a synchrotron beamline. In addition to being an effective alternative to a full pointwise scan to acquire a dark-field image of the sample, FAST also produces accurate quantitative measurements of its physical properties. For our live demonstration of a \qtyproduct{200 x 40}{points} with a measurement time of \qty{0.5}{\s\per point}, the FAST decision-making time was negligible, leading to an overall saving of \qty{\approx 80}{\minute} (about \qty{\approx 65}{\%}) of the experiment time. This saving was facilitated by our choice to acquire a batch of 50 measurements between the selection of the prospective measurement points. This ensured that the communication time stayed negligible with no noticeable loss in the quality of points acquired when compared to a pointwise candidate selection scheme (see Supplemental Materials, Fig. S1). 

The generalizability of the FAST method comes from the fact that the key NN-based component of this workflow is trained on just the standard cameraman image~\cite{cameraman}, not on close analogues of a sample of interest. While this generalizability results in a slight loss of performance of the technique , it still shows excellent sparsity performance for cases tested in previous research \cite{zhang_ei_2018,grosche_ieee_trans_comput_imaging__2020} and in the current work.  This has the benefit that we do not need \textit{a priori} knowledge of the sample. As such, while general pre-training would be difficult to satisfy for new and expensive experiments, the FAST approach can be used directly. Furthermore, the batch prediction and route optimization approach we implement can also be directly applied in any application of choice. Moreover, the experimental application of our work uses an extensible edge device and the widely used EPICS platform for hardware control, both of which can be incorporated into any instrument even with the SLADS-Net replaced by any other sampling strategies. For example, we could just replace the dark-field detection procedure described here with a fluorescence counting setup and use exactly the FAST scheme  for a fluorescence-based imaging of the sample. Alternatively, since all the instruments at the APS rely on EPICS controls, one can perform transmission, surface scattering, or any other 2D scanning experiment in any applicable beamline with only minor changes to the FAST routine.

The computations in the current workflow have a time complexity of \sloppy{$O\left(2N \log{N} + k M \log{N}\right)$}, where $N$ is the number of measured points, $M$ the number of unmeasured points, and $k$ the number of nearest neighboring measurements ($k=10$ in our case) that we use for the feature vector calculations. Here, the first term accounts for the creation  of the nearest neighbor K-d tree and the second term for the nearest neighbor calculation. The remainder of the algorithm has a linear time complexity and could be performed in parallel for the unmeasured points. We expect that it is possible to reduce this complexity using an approximate nearest neighbor search method instead of the K-d tree approach. As such, a GPU-based implementation that takes advantage of the parallelization and the approximation would likely significantly reduce the computation time. This stands in stark contrast with the time complexity of $O\left(N^3\right)$ (for $N$ measured points) for Gaussian Processes, a similarly training-free method that is widely used for autonomous experimentation. For an illustrative example,  Vasudevan \textit{et al}\cite{vasudevan_acsnano_2021} report a GP-based scanning microscopy experiment where the calculation of each set of measurement candidates takes \qty{\approx 6}{\s} on an NVIDIA DGX-2 GPU for a \qtyproduct{50 x 50}{} image; our workflow performs an equivalent calculation for a larger \qtyproduct{200 x 40}{} image within \qty{\approx 1.5}{\s} in a low-power CPU. We note, however, that GPs remain a very powerful and generalizable approach with a bevy of applications beyond only scanning microscopy.


We caution that our workflow suffers from three important challenges. First, it depends heavily on the initial $1\%$ random sampling to discover regions of heterogeneity. If an isolated feature present in an otherwise homogeneous region is not partially sampled during this random sampling step, then such a feature can be missed until much later in the scanning experiment (see \Cref{fig:experiment}H). A related second limitation is that this method produces sub-optimal reconstructions if the sample is sufficiently heterogeneous  that the data in each pixel changes significantly from pixel to pixel throughout the image (Supplemental Material in Hujsak et al~\cite{hujsak_micron_2018}). 
The third limitation, more practical in nature, is that the scan paths require significant motor movement, often including a retracing over points already measured.  As such, there could exist scenarios in which the time required for the motor movement eclipses the time required for a single measurement. We expect to address these limitations by explicitly including a measurement-density-based term ~\cite{grosche_ieee_trans_comput_imaging__2020} or a movement-time-based term in the candidate selection procedure~\cite{betterton_icra_2020}, or by using a line-based sampling technique \cite{helminiak_ei_2021}. 

Despite these challenges, we believe that the proposed FAST technique has great potential. It is an ideal tool for use 
cases with limited sampling or dosage budgets. It can be used to isolate regions of interest in sparse settings, to prepare for pointwise scanning in these regions. More generally, it can be used to guide any scanning microscopy experiment where we do not need full pointwise information. In the future, we expect to extend this method for 3D imaging, fly scans, ptychography, and other imaging applications. We expect that these developments will significantly enhance the efficacy of scanning microscopy experiments, bolstering their use for the study of dynamic physical phenomena. 

\section{Methods}\label{sec:methods}

\subsection{The SLADS-Net algorithm}\label{slads-net}

The SLADS-Net algorithm~\cite{zhang_ei_2018} used within the FAST workflow is an adaptation of the Supervised Learning Approach for Dynamic Sampling (SLADS) algorithm originally developed by Godaliyadda et al~\cite{godaliyadda_ei_2016}, and the algorithms differ only in their training approaches (~\Cref{subsec:training}). To explain the SLADS algorithm, we first denote the object we want to measure as $A\in\mathbb{R}^N$, where $N$ is the total number of pixels in the image. Further, we can denote the pixel at location $1\leq s \leq N$ as $a_s$ so that a measurement at the location $s$ extracts the value $a_s$;  each measurement is thus characterized by the pair $\left(s,\, a_s\right)$. After $k$ measurements, then, we get the  $k\times 2$ measurement vector
\begin{align}\label{eq:slads_measurement}
    Y^k = 
    \begin{bmatrix}
    s^1 && a_{s^1}\\
    s^2 && a_{s^2}\\
    \vdots\\
    s^k && a_{s^k}
    \end{bmatrix}
\end{align}
Using these $k$ measurements, then, we can reconstruct (e.g. via interpolation) an estimate $\hA^k$of the true object $A$. The difference between $A$ and $\hA^k$ is denoted as the \textit{distortion} $D(A, \hA^k)$ and can be calculated using any chosen metric. In the current work, we define $D(A, \hA^k)$ to be the L2 norm: 
\begin{align*}
    D(A, \hA^k) =  ||A - \hA^k||^2.
\end{align*}

Given the measurement $Y^k$ and the reconstruction $\hA^k$, a new measurement at any location $s$ will presumably reduce the distortion in the reconstruction. We can denote this reduction in distortion (RD) as 
\begin{align}\label{eq:reduction_distortion}
R^{k,s} =  D(A, \hA^k) - D(A, \hA^{k,s}) 
\end{align}
where $\hA^{k,s}$ is the reconstruction that includes the newly added measurement at $s$. The goal of the SLADS algorithm is then to identify the pixel location that would maximize this reduction in distortion:
\begin{align}\label{eq:goal_slads}
s^{k+1} = \operatorname*{argmax}_s \; R^{k,s}
\end{align}
Of course, since we cannot know the value of the measurement $a_{s}$ or the ground truth $A$, SLADS bases its selection on the conditional expectation of reduction in distortion (ERD), which is defined as:
\begin{align}\label{eq:erd}
\overline{R}^{k,s} = \mathbb{E}\left[R^{k,s} \big| Y^k\right] \quad \text{so that} \quad s^{k+1} = \operatorname*{argmax}_s\; \overline{R}^{k,s}.
\end{align}
The algorithm assumes that we can compute the ERD at $s$  based on just the measurement state $Y_k$ as
\begin{align}\label{eq:slads_training}
\overline{R}^{k,s} = g(v^{k,s})
\end{align}
where $v^{k,s}$ is a location-dependent feature vector calculated using the measurement state $Y_k$. The goal of the SLADS training procedure is to estimate the function $g$.

\subsection{Training}\label{subsec:training}
The training procedure for the SLADS/SLADS-Net algorithm is a supervised procedure in which we generate a large number of $(v^{k,s}, \overline{R}^{k,s})$ pairs and use these to estimate $g$. Note that this is a pixelwise computation that is performed independently for each measurement location $s$; for each measurement $s$ we have to calculate a reconstruction $\hA^{k,s}$ before we can calculate the RD $R^{k,s}$. To make this computationally tractable, the Godaliyadda \textit{et al} \cite{godaliyadda_ei_2016} use approximations that ensure that the RD of each pixel only depends on its local neighborhood. Correspondingly, instead of working with the full measurement state $Y^k$, the training procedure uses carefully designed feature vectors that capture the local neighborhood of the pixel at location $s$. As shown in \Cref{fig:workflow}B, the feature vector for the pixel $P$ consists of six features: (i) $\nabla_x$ and $\nabla_y$ are the spatial gradients at $P$, (ii) $\sigma_{1,r}$ and $\sigma_{2,r}$ measure the deviation of the estimated value for $P$ from the nearby measured values (highlighted in red), and (iii) $L$ (which is the distance of $P$ from the closest measured point) and $\rho_r$ measure the density of measurements around $P$.

The original SLADS algorithm assumes that this feature vector is linearly related to the RD, and the training therefore is a linear regression procedure. The SLADS-Net adaptation first uses an radial basis function (RBF) kernelization to transform the 6-dimensional feature vector to a 50-dimensional vector, then replaces the linear predictor with a nonlinear fully-connected neural network that contains 5 hidden layers with 50 nodes each.

In this work, we train the SLADS-Net neural network on \textit{only} the standard cameraman image, without using any \textit{a priori} information about the sample. For the training, we generate a measurement state $Y^k$ by randomly choosing a fixed number number of measurement locations, then calculate the feature vector $v^{k,s}$ and the RD $\overline{R}^{k,s}$ for each unmeasured pixel. We generate such sets of training pairs for 10 different sample coverage percentages
between 1\% and 80\%. This overall comprises our training dataset. We use this data to train the neural network for 100 epochs using the Adam optimizer with the learning rate $0.001$. We use this trained model for all the simulated and experimental measurements. We provide an example of a training measurement set---the measured points, the interpolated reconstruction, and the corresponding RD for the unmeasured points---in the supplemental materials (Fig. S2)

\subsection{Experimental measurements}\label{subsec:exp_analysis}
At each point of the measurement, a tight region of interest (RoI) around the expected position of the thin film Bragg peak was extracted from the corresponding diffraction image. Integrated intensities of the RoI were used to guide the NN prediction. For the flat region, the integrated intensity is high, showing up as brighter contrast on the dark field image. For the deformed region, the integrated intensity is low (darker contrast on the dark field image) as the illuminated film diffraction partially exits the selected RoI (see Supplemental Materials, Fig. S3).

For the FAST experiment, the predicted ERD and the dark-field reconstruction served as visual guides to inform when to stop the experiment.. During the experiment, we noted that the ERD and the reconstruction had stabilized 
by \qty{\approx 20}{\%} scan coverage, but we let the experiment run to \qty{\approx 35}{\%} coverage to ensure that this behavior persisted (see  Supplemental Materials, Fig. S4). While we used this visual criterion for our exploratory experiment, it is straightforward to design a numerical stopping criterion based on the absolute or relative convergence of the ERD, or on the per-iteration change in the reconstructed image.

\section*{Data and Code Availability} 
The data and code will be made available at \url{https://github.com/saugatkandel/fast_smart_scanning}

\section*{Acknowledgments}
Work performed at the Center for Nanoscale Materials and Advanced Photon Source, both U.S. Department of Energy Office of Science User Facilities, was supported by the U.S. DOE, Office of Basic Energy Sciences, under Contract No. DE-AC02-06CH11357. {\color{red}} We also acknowledge support from Argonne LDRD 2021-0090 – AutoPtycho: Autonomous, Sparse-sampled Ptychographic Imaging. We gratefully acknowledge the computing resources provided on Bebop, a high-performance computing cluster operated by the Laboratory Computing Resource Center at Argonne National Laboratory. X.L. acknowledges support from the National Science Foundation CBET Program under the award no. 2025214.

\section*{Competing interests} 
The authors declare that they have no competing financial interests.

\bibliographystyle{naturemag}
\bibliography{reference_2,cd_refs_2,Tao}

\end{document}


\title{Supplemental material}
\maketitle

\setcounter{figure}{0}
\renewcommand{\thefigure}{S\arabic{figure}}%

\begin{figure}[ht]
	\includegraphics[width=0.99\textwidth]{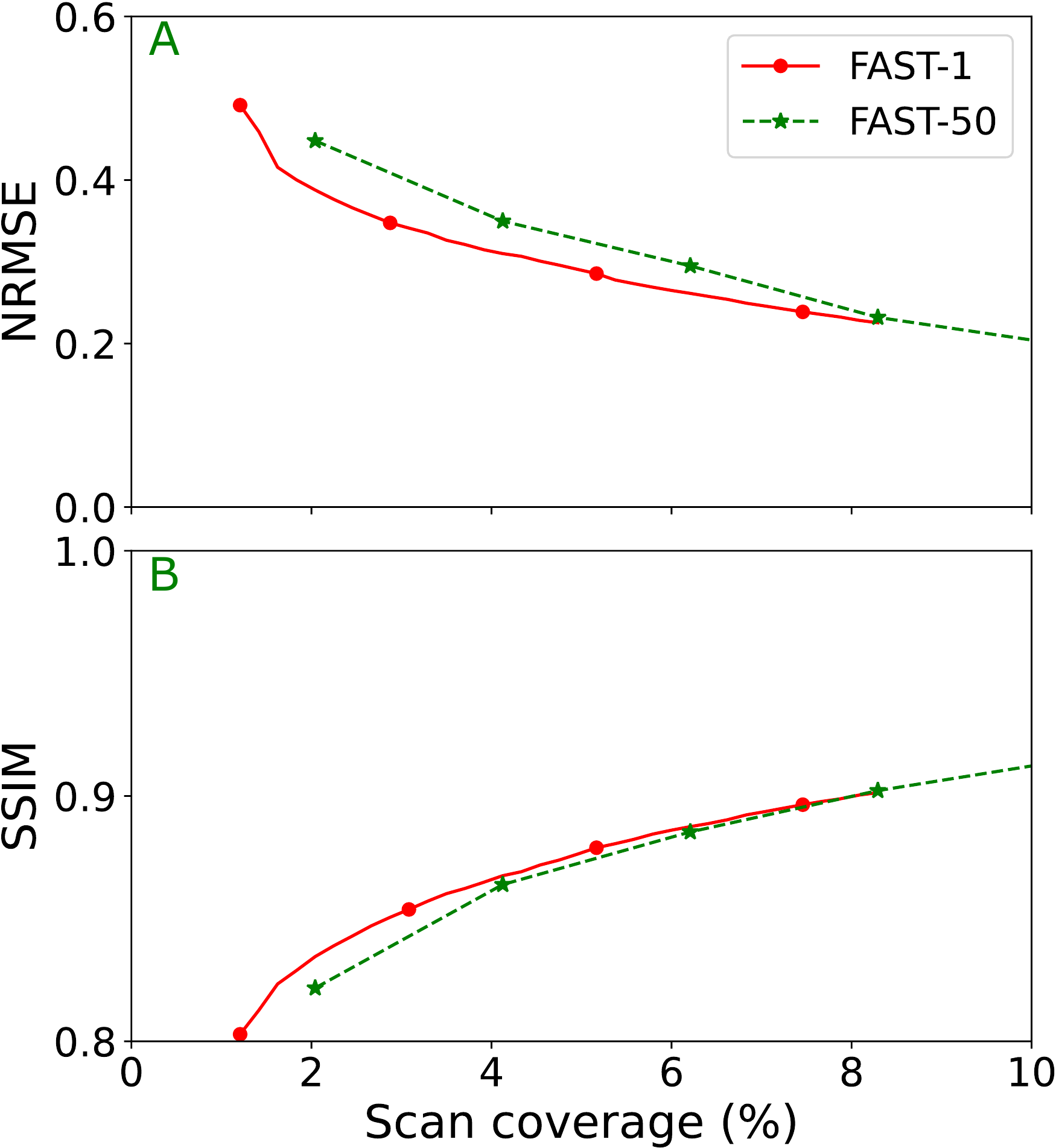}
	\caption{Comparison of the FAST reconstructions for scan batch size of 1 (FAST-1) and 50 (FAST-50) as a function of the scan coverage for the numerical simulation described in Section II.B. We observe that FAST-1 initially performs better, with lower NRMSE and higher SSIM, than FAST-50, but this advantage erodes quickly. We ended the FAST-1 experiment at \qty{\approx 8.2}{\%} sampling.due due to simulation time limitations.}\label{fig:supp_batch_comparisons}
\end{figure}
\begin{figure}[ht]
	\includegraphics[width=0.99\textwidth]{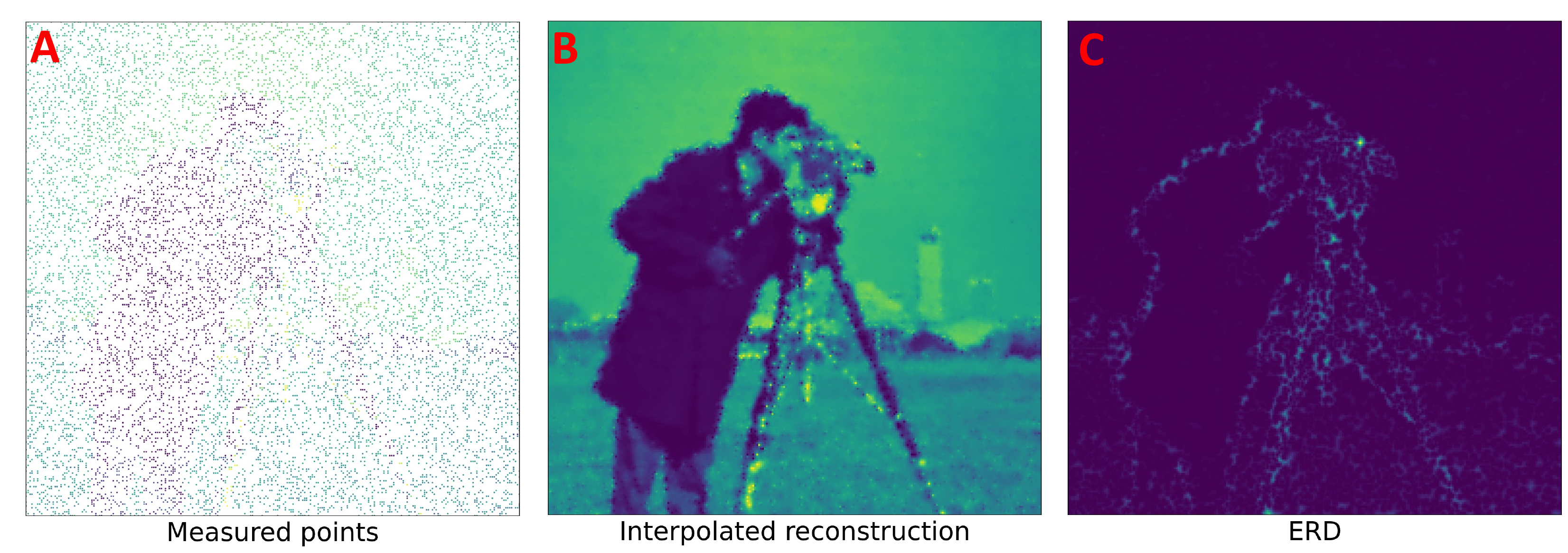}
	\caption{Example of training data. (A) shows a set of randomly selected measurement points. (B)shows the reconstruction calculated by interpolating these measurements. (c) shows the ERDs calculated for the unmeasured points, with the ERD highest at regions of hetereogeneity. The location of the measured points, the measured values, and the reconstruction are used to generate feature vectors for the training, and the ERDs are used as the training labels.}\label{fig:supp_example_training}
\end{figure}
\begin{figure}[ht]
\centering
	\includegraphics[width=0.99\textwidth]{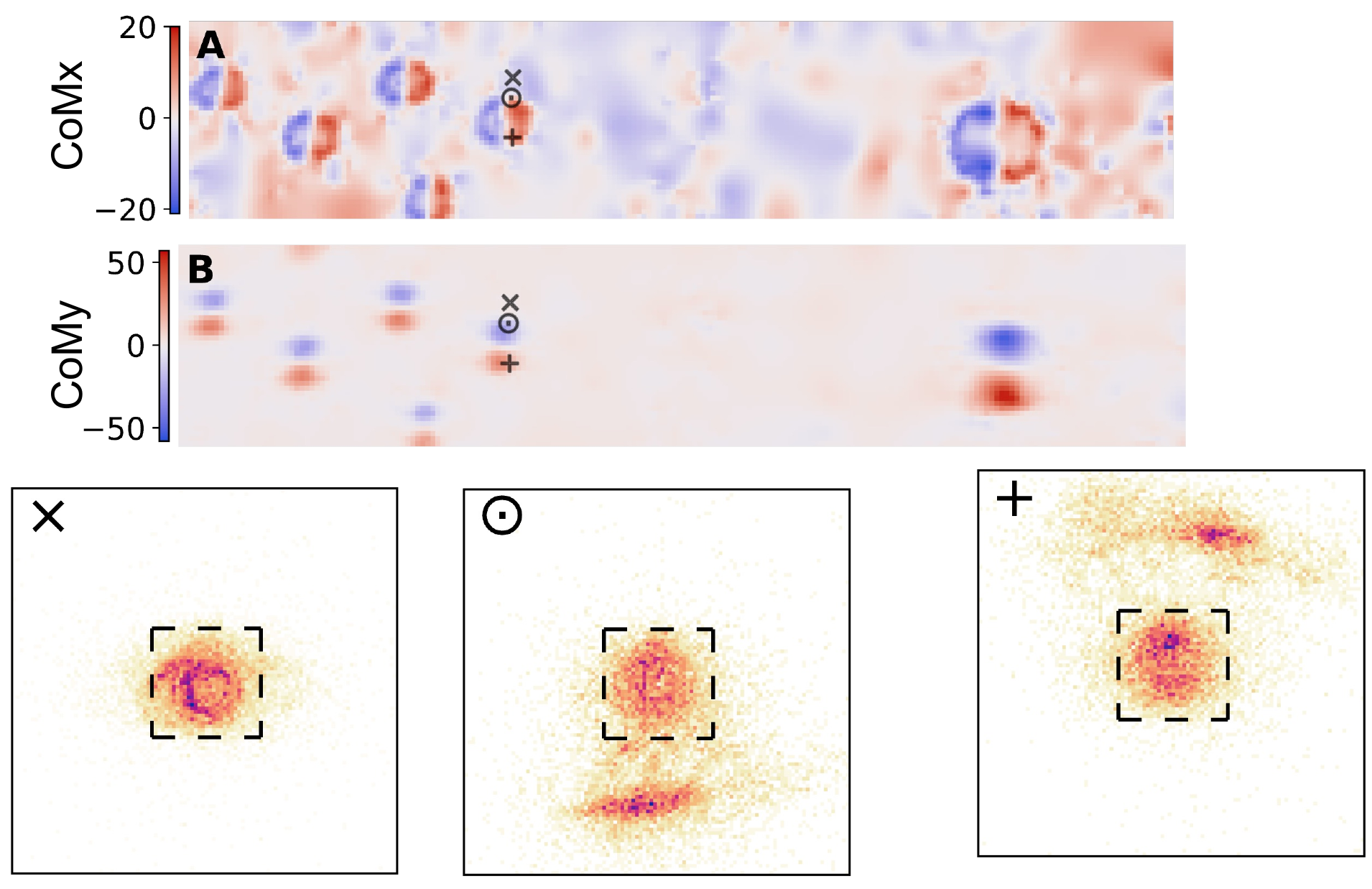}
	\caption{Example of ROI selection and change in diffraction patterns around the bubbles. (A) and (B) respectively show the CoMx and CoMy calcualted from the FAST scan with 20\% covergae, as discussed in Section II.C. The diffraction patterns for the points marked with the $\times$, $\bigodot$, and $+$ are shown in the bottom row. The $\times$ point is in a region without a bubble and has the diffraction pattern at the Bragg angle. The points marked with $\bigodot$ and $+$ are located at the top and bottom edges of the bubble, and therefore show additional anomalous diffraction spots. The dashed square boxes in the diffraction pattern figures indicate the ROI used for the dark-field image reconstructions (shown in Figure 4 in the main paper). The CoM calculations use the regions outside the dashed square boxes as the RoI.}\label{fig:supp_example_diffs}
\end{figure}
\begin{figure}[ht]
\centering
	\includegraphics[width=0.99\textwidth]{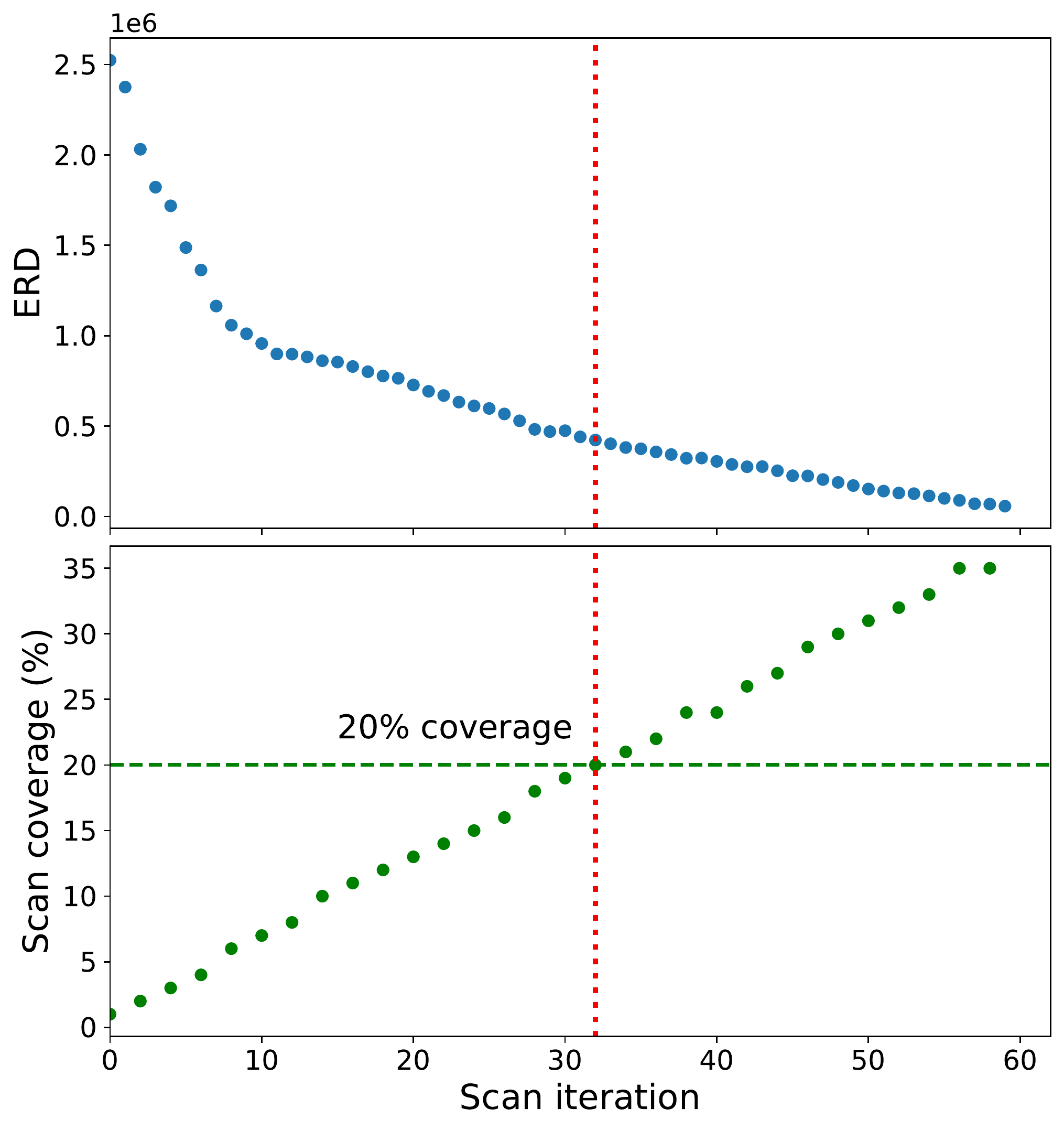}
	\caption{Evolution in the ERD for the experimental demonstration. The ERD initially decreases rapidly, during which point the each batch of 50 points significantly improves the sample reconstruction. At per-iteration change in the ERD is much smaller at 20\% coverage.}\label{fig:supp_erds}
\end{figure}